\documentclass[preprint,12pt]{elsarticle}
\usepackage{amsmath,amsfonts,amssymb,mathbbol,dsfont}
\usepackage{graphicx,psfrag,color}
\usepackage{dcolumn}
\usepackage{bm}
\usepackage{mathbbol, appendix}

\def\dim{{\sf dim}_{\mathbb{C}}}
\def\dual{\ \stackrel{\Phi_\d}{\longrightarrow}\ }

\def\d{{{\sf d}}}
\def\r{{\bm{r}}}
\def\x{{\bm{x}}}

\def\one{{\mathbb{1}}}

\def\im{{\sf {i}}}

\journal{Nuclear Physics B}

\makeindex
\begin{document}
\begin{frontmatter}

\title{Fock Parafermions and Self-Dual Representations of the Braid Group}

\author[]{Emilio Cobanera$^{a,}$\footnote{
Corresponding author: cobanera@lorentz.leidenuniv.nl}
and Gerardo Ortiz$^b$}

\address{$^a$Instituut-Lorentz, Universiteit Leiden, P.O. Box 9506, 2300
RA Leiden,  The Netherlands,\\
$^b$Department of Physics, Indiana University, Bloomington,
IM 47405, USA.}

\begin{abstract}
We introduce and describe in second quantization a family of particle species
with \(p=2,3,\dots\) exclusion and \(\theta=2\pi/p\) exchange statistics.
We call these anyons {\it Fock parafermions}, because they
are the {\it particles} naturally associated to the parafermionic zero-energy {\it modes},
potentially realizable in mesoscopic arrays of fractional topological insulators.
Their second-quantization description entails the concept of {\it Fock algebra}, i.e.,
a Fock space endowed with a
statistical multiplication that captures and logically correlates
these anyons' exclusion and exchange statistics. As a consequence
normal-ordering continues to be a well-defined operation. Because of its relevance to
topological quantum information processing, we also derive families of self-dual
representations of the braid group for any $p$, with the Gaussian representation 
being a special case. The self-dual representations can be realized in terms of {\it local} quadratic combinations
of either parafermions or Fock parafermions, an important requisite for physical
implementation of quantum logic gates.
\end{abstract}

\end{frontmatter}

\section{Introduction: The problem of anyons in second quantization}

During the last decade quantum information science has evolved
into a mature discipline, with challenging theoretical developments actually 
resulting in intense experimental work.  Large scale quantum information 
processing devices may not become readily available anytime soon, but 
when they do become a reality, they will offer enormous advantages to scientists and engineers.  
The challenge, or hope rather, that constructing the inner gears of physical reality 
in the right way may foster quantum information processing is so great that 
it is forcing us to reconsider fundamental principles.
 Consider the fundamental aspects of quantum mechanics in two space dimensions. 
 Topological quantum computation exploits braiding at the hardware level to 
 minimize the environment's decohering effects \cite{pachos}, thus constituting 
 an example of utilizing fundamental physical principles as engineering assets.

In this paper we are concerned with the Fock space representation and
braiding of anyons, which are indistinguishable particles with unconventional statistics.   
We call the second quantized description of a particular type of anyon a Fock parafermion. 
Fock parafermions are tied to localized excitations of clock
models \cite{pclock,conII},  edge modes of quantum Hall liquids \cite{lindner}, and 
fractional topological insulators in general. In this paper we introduce the concept 
of self-dual representation of the braid group \cite{turaev} and obtain an extended 
family of  examples. Before describing the specific contents and main results of
this paper, let us briefly review the concept of fractional exchange statistics
as developed in the framework of first quantization.

The non-relativistic (first quantized) description 
of $N$ indistinguishable particles, each constrained to a $d$-dimensional 
coordinate space $\mathds{R}^d$,
relies on the analysis of the topological 
properties of  the system's classical {\it physical} configuration space 
${\cal C}_N(d)$ \cite{laidlaw1971,leinaas1977}. 
The latter is defined as a modification of 
the usual Cartesian product $\mathds{R}^{dN}$ for distinguishable particles. 
This modification entails removing the points where two or more particles coincide, 
and then identifying points that are equal up to a permutation of single-particle indices.  
Hence each physical configuration of the $N$ indistinguishable
particles corresponds to {\it a unique} point in ${\cal C}_N(d)$. 
The topological properties of ${\cal C}_N(d)$ are characterized 
by the homotopy equivalences among physical trajectories 
that effectively exchange two particle indices. The particles' statistical exchange 
properties should be independent of their trajectories; two trajectories that can be continuously 
deformed to each other are equivalent. As it turns out,
the first homotopy group of the configuration space is the permutation group 
of $N$ objects, $S_N$, whenever $d\geq 3$, and it is the braid group $B_{N}$ 
of $N$ strands for $d=2$  \cite{leinaas1977}. This implies that in higher space 
dimensions particles may be either fermions or bosons while in $d=2$, there 
are possibilities for more exotic particles, such as anyons, which show fractional statistics.

It is important to emphasize that this way of characterizing particle statistics 
is {\it kinematical}, meaning that it does not  invoke any 
particular Hamiltonian. Indeed, one can 
define exchange statistics as a phenomenon of parallel 
transport (without Hamiltonians) and the statistical phase as the holomony 
associated with the geometric evolution leading to the exchange path, i.e., 
a statistical evolution. 
This is analogous to the definition of Berry phases and its 
generalizations \cite{shapere}, which are holonomies, associated with adiabatic 
parallel transport over a generic parameter space. Based on the character of the 
physical state, which must belong to a Hilbert space, one can realize statistical evolutions 
which may be Abelian or non-Abelian, depending on the dimension of the unitary 
matrix representing the holonomy. Thus, one may speak of non-Abelian exchange 
statistics whenever that matrix is at least two-dimensional while the simple 
scalar case will be reserved for Abelian statistics.

Only multi-component states allow for non-Abelian statistical evolutions. 
In practice one can simulate a multi-component state by choosing a subspace 
degenerate in energy or some other conserved quantity, such as the ground 
state of a Hamiltonian. Then one can perform an adiabatic evolution to
realize the desired non-Abelian exchange.
 The implementation of non-Abelian statistical phases is {\it dynamical} and 
 subject to errors, including departures from adiabaticity. This particular 
 type of error can be systematically accounted for by using an adiabatic 
 perturbation theory \cite{rigolin}. Holonomic quantum 
computation \cite{pachos} takes this observation and extends its scope 
by exploiting adiabatic transport in arbitrary non-Abelian unitary vector
bundles to achieve quantum logic gates. Topological quantum computation \cite{pachos} 
is a particular instance of holonomic computation where the transported subspace of 
degenerate states has special physical properties. 
Typically, it is the manifold of ground states for {\it topological quantum matter} 
that are said to be {\it topologically protected} or robust against local 
perturbations. The aim of topological quantum computation is to exploit this 
fact at the hardware level for fault tolerant quantum information processing.  

Topological quantum matter represents many-body systems whose degenerate 
ground states cannot be distinguished by local measurements in the bulk \cite{zohar}. 
They typically  support 
{\it localized} modes or excitations that may be carefully manipulated to 
effectively exhibit anyonic exchange statistics. For example 
\cite{carlo,hyart2013,ivanov}, it is common to hear that Majorana modes 
are non-Abelian anyons, because they support a non-Abelian, unitary 
representation of the braid group.  However, Majoranas are not particles. By 
particle, we mean the usual quantum field theory 
notion, which requires a Fock space and thus, second quantization; 
that is, one needs a way of {\it counting} them. That said, by generalizing the 
concept of particle statistics to  
localized excitations of topologically ordered vacua, we acquire a new source 
of robustness. These localized excitations support natural representations of 
the braid group and are optimal candidates for  
the realization of the elusive topological quantum computer. 

In this paper, we study a generalization
of Majorana modes called {\it Weyl para\-fermions} \cite{Note}. 
Parafermions are effective degrees
of freedom or {\it modes}  associated to clock models \cite{fradkin1980,alcaraz1981} that,
as we will show, support a variety of non-Abelian representations of the braid group
(including the Majorana representation as a special case).  Moreover, 
they are naturally associated to {\it particle} 
degrees of freedom that we call {\it Fock parafermions}. Fock parafermions 
are indistinguishable particles 
displaying unconventionally correlated $p$-exclusion and $\frac{2\pi}{p}$-exchange statistics, 
where $p$ is an integer larger than one. Most importantly, we 
introduce self-dual representations of the braid group for all $p$ which can 
be realized in terms of quadratic combinations 
of either Weyl or Fock parafermions, a crucial asset for physical implementation. 

The outline of the paper follows. In Section \ref{Section2} we present the 
Weyl algebra with generators $U$ and $V$ originally introduced to study 
finite-dimensional quantum 
mechanics. There we show a fundamental canonical decomposition of the algebra 
in terms of Weyl algebras with smaller periods. This technical result is 
used to analyze the self-dual representations of the braid group developed 
in later sections. It turns out that the Weyl algebra is behind the realization 
of Weyl parafermions, or simply parafermions, with Majoranas being identified 
with their lower dimensional representation $p=2$. This is the topic of 
Section \ref{Weyl_parafermions}, where we derive parafermions
 as the product of {\it order} and {\it disorder} variables associated with 
 self-dual quantum clock models \cite{conII}. Next we present 
the self-dual representations of the braid group in Section \ref{sdreps_section}, 
which is one of the main results of the paper,  and we 
show how to realize these representations in terms of parafermions.  We also show that 
the Gaussian representation of 
Jones \cite{j,gj} is a particular example of a self-dual representation.  Our next main
result is unveiled in Section \ref{Fock_parafermions}, where we 
show how to construct a Fock space for the Fock parafermions 
that reduce to canonical fermions in the case $p=2$.  The key
technical development is the introduction of the concept of 
{\it Fock algebra} with a {\it statistical} multiplication. The 
advantage of this approach
is that it allows us to {\it derive} the unconventional algebra
of creation and destruction operators of Fock parafermions. It is possible that
this approach may be generalized to obtain Fock spaces for anyon-like particles
with more complicated statistical properties.
Majoranas are mathematically realized as combinations of fermionic creation
and annihilation operators. Fock parafermions possess an unconventional 
algebra of creation and annihilation operators which allows us to generalize 
this connection to all Weyl parafermions, as shown in Section \ref{cs_to_gs}. 
Finally, \ref{whcb} describes a Jordan-Wigner-like relation between 
Fock parafermions and a new type of particle that we call Weyl hard-core bosons, 
which have a Fock space representation with $p$-exclusion and bosonic statistics. 

\section{Finite-dimensional quantum mechanics: Weyl algebra}
\label{Section2}

Heisenberg's canonical commutation relations for position, $X_i$, and momentum, 
$P_i$, Hermitian operators
\begin{equation}\label{heisenberg}
[X_i,P_j]=\im \hbar\delta_{i,j}\,\mathds{1} , \qquad i,j=1,\dots,M ,
\end{equation}
cannot be satisfied by finite-dimensional matrix representations \cite{norm}. 
This is unfortunate because it is natural to seek insight
into quantum mechanics by disentangling 
conceptual issues from the mathematical complications associated to
infinite-dimensional state spaces. To realize a finite-dimensional 
framework for quantum mechanics, Hermann Weyl \cite{weyl,schwinger} 
proposed a different 
starting point: the algebra  \(W(M)\)
generated by translations in position and momentum spaces,
\begin{equation}
V_{i,{\sf x}}\equiv e^{\im {\sf x} P_i/\hbar},\quad U_{j,{\sf p}}\equiv 
e^{\im {\sf p}X_j/\hbar}, \qquad i,j=1,\dots,M.
\end{equation}
These unitary generators commute for \(i\neq j\), and otherwise satisfy
\begin{equation}
V_{i,{\sf x}}U_{i,{\sf p}}=e^{\im {\sf x p}/\hbar}U_{i,{\sf p}}V_{i,{\sf x}}. 
\end{equation}

The Weyl algebra \(W(M)\) does admit a natural finite-dimensional truncation. 
Let us introduce  a lattice (fundamental) spacing \(\delta\), and discretize \(\sf x, p\) as
\begin{equation}
{\sf x}_m=m \, \delta, \quad {\sf p}_n=\frac{2\pi \hbar n}{p \, \delta}, 
\qquad m,n\in\mathds{Z},
\end{equation}
with \(p=2,3,\dots\) a fixed positive integer. Then, we can define
\begin{equation}
V_i\equiv e^{\im \delta P_i/\hbar} , \qquad U_j\equiv e^{\im 2\pi X_j/p\delta} , 
\end{equation}
so that \(V_{i,{\sf x}_n}=V_i^n\), \(U_{j,{\sf p}_m}=U_j^m\), and 
\begin{equation}\label{usandvs}
V_iU_i=\omega U_iV_i, \qquad \omega\equiv e^{\im 2\pi/p}, 
\qquad \bar{\omega}=e^{-\im 2\pi/p}. 
\end{equation}
It was Weyl who noticed \cite{weyl} that the relations of 
Eq. \eqref{usandvs} {\it do} admit a finite dimensional, 
unitary representation \(W_p(M)\) in a Hilbert space (defined over 
the field of complex numbers $\mathbb{C}$) of dimension \(p^M\).
In particular, the unitary \(p\times p\) matrices 
\begin{eqnarray} \hspace*{-0.5cm}
{V}=
\begin{pmatrix}
0& 1& 0& \cdots& 0\\
0& 0& 1& \cdots& 0\\
0& 0& 0& \cdots& 0\\
\vdots& \vdots& \vdots&      & \vdots \\
0& 0& 0& \cdots& 1\\
1& 0& 0& \cdots& 0\\
\end{pmatrix}, \qquad
{U}=\begin{pmatrix}
1& 0& 0& \cdots& 0\\
0&\omega & 0& \cdots& 0\\
0& 0&\omega^2 & \cdots& 0\\
\vdots& \vdots& \vdots&      & \vdots \\
0& 0& 0& \cdots& 0\\
0& 0& 0& \cdots&\omega^{p-1} \\
\end{pmatrix} .  
\label{VsandUs}
\end{eqnarray}
satisfy \(VU=\omega UV\), and have finite period \(p\) 
(\(V^p=\mathds{1}=U^p\)). Hence the {\it Weyl generators}
\begin{eqnarray}
V_i&\equiv& \mathds{1}\otimes\dots\otimes V\otimes\dots \otimes 
\mathds{1}, \quad (\mbox{\(i\)th
position}), \quad i=1,\dots,M ,\\
U_j&\equiv& \mathds{1}\otimes\dots\otimes U\otimes\dots \otimes 
\mathds{1}, \quad (\mbox{\(j\)th position}),\quad j=1,\dots,M ,
\end{eqnarray}
afford the desired finite-dimensional, unitary representation of Eq. \eqref{usandvs}. 
For \(p=2\), \(\omega=-1\), \(V=\sigma^x\), and \(U=\sigma^z\) (where 
$\sigma^{x,z}$ are Pauli matrices). It is interesting to notice that while 
Planck's constant explicitly appears in $W(M)$, 
it does not appear in \(W_p(M)\), and so there is no obvious notion of 
classical, $\hbar \rightarrow 0$, limit for this form of discrete quantum mechanics.

It is well known that the three Pauli matrices 
\(\sigma^x,\sigma^z,\sigma^y=\im \sigma^x\sigma^z\)
close a Lie algebra. There is an interesting generalization of this
fact to arbitrary \(p\). Let ${\bf m}=(m_1,m_2)$ with $m_{1,2}=0,\dots,p-1$, 
non-negative integers. Define the unitary Schwinger basis
\begin{equation}
J_{\bf m}\equiv \bar{\omega}^{m_1m_2/2}V^{m_1}U^{m_2}.
\end{equation}
The set $\{J_{\bf m}\}$ realizes 
an orthogonal (with respect to the Hermitian inner product defined by the trace) 
basis of the full matrix algebra \(M_p(\mathds{C})\) of \(p\times p\) complex 
matrices. Moreover, the Schwinger basis is closed under the bracket operation,
\begin{equation}
[J_{\bf m}, J_{\bf n}]= 2\im \sin\Big(\frac{\pi}{p}({\bf m}\times {\bf n})\Big) 
J_{{\bf m}+{\bf n}}\ .
\end{equation}
Throughout this paper, addition and multiplication of integers
are understood to be modulo \(p\) operations unless stated otherwise.
Discarding the identity \(J_{(0,0)}=\one\),
we recognize this to be the Lie algebra of \(p\)-dimensional complex, 
traceless matrices $su(p-1)$ \cite{fairlie1990}.

The Weyl algebra  $W_p(M)$ finds one of its most important applications in the 
design of model Hamiltonians with unconventional phase 
transitions \cite{pclock,conII}, non-commutative symplectic geometry 
\cite{aldrovandi,wess}, measurement-assisted topological quantum computation
\cite{lindner}, and the study of interacting zero-energy edge 
modes \cite{fendley}. Historically, 't Hooft's work on confinement in QCD
was instrumental to popularizing Weyl's algebra in the late 1970s.
In Ref. \cite{thooft}, he proposed a generalized order parameter
for \(SU(N)\) gauge theories satisfying the algebra of the Weyl 
generators. A flurry of activity  followed, centered on studying quantum many-body 
systems with Weyl generators as elementary degrees of freedom. 
Popular examples are the vector Potts, or $p$-clock, 
model for the BKT phase transition \cite{pclock},
\(\mathds{Z}_p\)  gauge theories for 
confinement transitions \cite{weinstein}, and, more recently,
topologically  ordered \(\mathds{Z}_p\) toric code models \cite{zptc}.

In closing this section on Weyl algebras, we want to show 
that the Weyl algebra \(W_p\) 
admits a canonical decomposition in terms of 
 Weyl algebras $W_{p_j}$ with periods $p_j$, 
\begin{equation}\label{tensor_decomp}
W_{p_1}\otimes\dots \otimes W_{p_t}\cong W_p \ ,
\end{equation}
dictated by the decomposition of \(p\) into relative prime factors,
\begin{equation}\label{p_decomp}
p=\prod_{j=1}^{t} p_j,\qquad p_j=q_j^{m_j},\qquad q_j\ \ 
\mbox{distinct prime numbers,}\ \ j=1,\dots,t ,
\end{equation}
with $m_j$ and $t$ integers. The isomorphic mapping \(\cong\) will be very useful  
for establishing some basic properties 
of the self-dual representations of the braid group
in Section \ref{sdreps_section}.  

Assume \(p=p_1p_2\), with \(p_1\) and \(p_2\) 
relative primes. Then the mapping 
\begin{eqnarray} \label{cong_otimes1}
U_{(p_1)}^m\otimes U_{(p_2)}^m\cong U_{(p)}^m,\quad
V_{(p_1)}^m\otimes V_{(p_2)}^m\cong V_{(p)}^{(p_1+p_2)m},\qquad m=0,\dots, p-1, 
\end{eqnarray}
induces the algebra isomorphism 
\begin{equation}
W_{p_1}\otimes W_{p_2} \cong W_p \ .
\end{equation}
(Here we write \(U_{(p_j)},\ V_{(p_j)}\) for the Weyl 
generators of \(W_{p_j}\) instead of 
 \(U,\ V\), and \(\omega_{(p)}\) instead of  \(\omega\). 
For example, \(\omega_{(2)}=-1\),
\(U_{(2)}=\sigma^z\) and \(V_{(2)}=\sigma^x\).)
To see that \(\cong\) is an algebra isomorphism we need to combine
two observations. First, the two relations
\begin{eqnarray}
(V_{(p_1)}^m\otimes V_{(p_2)}^m)(U_{(p_1)}^n\otimes U_{(p_2)}^n)\!\!&=&\!\!
\omega_{(p_1)}^{mn}\omega_{(p_2)}^{mn} (U_{(p_1)}^n\otimes U_{(p_2)}^n)
(V_{(p_1)}^m\otimes V_{(p_2)}^m),\\
V_{(p)}^{(p_1+p_2)m}U_{(p)}^n\!\!&=&\!\! \omega_{p}^{(p_1+p_2)mn}U_{(p)}^nV_{(p)}^{(p_1+p_2)m},
\end{eqnarray}
are satisfied because
\begin{equation}
\omega_{(p_1)}^{mn}\omega_{(p_2)}^{mn}=\omega_{(p)}^{(p_1+p_2)mn}\ .
\end{equation}
Second, the mapping
\begin{equation}
m\mapsto (p_1+p_2)m\ \ ({\sf mod}\ p),\qquad m=0,\dots, p-1,
\end{equation}
is one-to-one and onto, but {\it only if} \(p_1,p_2\) are relative primes. 
The isomorphism fails to be one-to-one if one does not enforce this latter condition. 
For example, blindly applied to the case \(p_1=2,p_2=2,p=4\), 
Eq. \eqref{cong_otimes1} obtains 
\begin{equation}
\sigma^x\otimes \sigma^x \cong V_{(4)}^4=\mathds{1}\ ,
\end{equation}
mapping an element that is not the identity to the identity.
Since by definition the factors \(p_j\) in Eq. \eqref{p_decomp} are 
relative primes, repeated application of Eq. \eqref{cong_otimes1} establishes
the decomposition of Eq. \eqref{tensor_decomp}.

Consider as an example the decomposition \(W_2\otimes W_3\cong W_6\): 
\begin{eqnarray}
\sigma^z\otimes U_{(3)}&\cong & U_{(6)}\ , \qquad 
\ \sigma^x\otimes V_{(3)}\ \cong\  V_{(6)}^5\ ,\nonumber\\
\mathds{1}_{(2)}\otimes U_{(3)}^2&\cong& U_{(6)}^2\ , \qquad 
\mathds{1}_{(2)}\otimes V_{(3)}^2\, \cong\ V_{(6)}^4\ ,\nonumber\\
\sigma^z\otimes \mathds{1}_{(3)}&\cong& U_{(6)}^3\ , \qquad \ 
\sigma^x\otimes \mathds{1}_{(3)}\ \cong\ V_{(6)}^3\ ,\\
\mathds{1}_{(2)}\otimes U_{(3)}&\cong& U_{(6)}^4\ , \qquad
\mathds{1}_{(2)}\otimes V_{(3)}\, \cong\ V_{(6)}^2\ ,\nonumber\\
\sigma^z\otimes U_{(3)}^2&\cong& U_{(6)}^5, \qquad \ \
\sigma^x\otimes V_{(3)}^2\ \cong\ V_{(6)}\ .\nonumber
\end{eqnarray}

The reasoning leading to \(\cong\) can be adapted to obtain 
canonical embeddings \(W_q\hookrightarrow W_p\)
of lower into higher dimensional Weyl algebras. If 
\(p/q\) is an integer and \(p/q\) and \(q\) are relative primes,
then the mapping 
\begin{eqnarray}\label{canonical_embedding} 
U_{(q)}^{mp/q}\hookrightarrow U_{(p)}^{mp/q},\quad 
V_{(q)}^{mp/q}\hookrightarrow V_{(p)}^{mp^2/q^2},\qquad m=0,\dots, q-1, 
\end{eqnarray}
defines a one-to-one homomorphism. We can read off our example in the previous
paragraph the embeddings \(W_2\hookrightarrow W_6\) and \(W_3\hookrightarrow W_6\), 
\begin{eqnarray}
\sigma^z&\hookrightarrow & U_{(6)}^3\ , \qquad 
\ \sigma^x\ \hookrightarrow\  V_{(6)}^3\ ,\\
\ \nonumber\\
U_{(3)}&\hookrightarrow& U_{(6)}^4\ , \qquad
V_{(3)}\ \hookrightarrow\ V_{(6)}^2\ ,\\ 
U_{(3)}^2&\hookrightarrow& U_{(6)}^2\ , \qquad 
V_{(3)}^2\ \hookrightarrow\ V_{(6)}^4\ .
\end{eqnarray}
Similarly to the isomorphism \(\cong\), the embeddings \(\hookrightarrow\) 
will be very useful
to understand some properties of our self-dual representations of the braid
group in Section \ref{sdreps_section}.
 
\section{Weyl parafermions}
\label{Weyl_parafermions}

In this section we want to illustrate the natural relation between 
Majorana fermions and the finite-dimensional, $p=2$, representation of 
the Weyl algebra defined in 
previous section. This fact allows a generalization of the Majorana concept 
to higher-dimensional representations, $p>2$, in terms of quantum 
degrees of freedom called Weyl parafermions or, simply, parafermions. 

The operators
\begin{eqnarray}
\Gamma_i\equiv V_i \Big(\prod_{j=1}^{i-1} U_j\Big)\ ,\qquad 
\Delta_i\equiv V_i U_i \Big(\prod_{j=1}^{i-1} U_j\Big)=\Gamma_i U_i\ ,
\label{uvs2pfs}
\end{eqnarray}
define an alternative set of generators of the Weyl algebra \(W_p(M)\), as follows
from the inverse relations
\begin{equation}
U_i=\Gamma_i^\dagger \Delta_i^{\;},\qquad 
V_i=\Gamma_i^{\;}\Big(\prod_{j=1}^{i-1} \Delta_j^\dagger \Gamma_j^{\;}\Big)\ .
\label{gd2uvs}
\end{equation}
The new generators satisfy 
\begin{equation}\label{omegacommutation}
\Gamma_{i}\Gamma_{j}=\omega \Gamma_{j}
\Gamma_{i},\quad \Delta_i\Delta_j=\omega \Delta_j\Delta_i,
\qquad \mbox{for}\quad i<j,
\end{equation}
\begin{equation}\label{omegacommutation2}
\Gamma_i\Delta_j=\omega\Delta_j\Gamma_i,\qquad 
\mbox{for}\quad i\leq j,
\end{equation}
and 
\begin{equation}\label{power}
\Gamma_{i}^p=\mathds{1}=\Delta_i^p,\qquad \Gamma_{i}^{p-1}=\Gamma_i^\dagger, \qquad
\Delta_i^{p-1}=\Delta_i^\dagger\ .
\end{equation}
Regarded as quantum degrees of freedom, the \(\Gamma_i,\ \Delta_i\) are collectively 
called {\it parafermions} \cite{alcaraz1981,conII}. 
We will come back to the physical meaning and the context in which 
parafermionic excitations emerge later in this section. 

In the mathematical literature, the algebra realized by parafermions \cite{Note1}
is known as a generalized Clifford algebra \cite{kwasniewski,smith,traubenberg}.
This is in part because parafermions generate
the standard Clifford algebra for the special case \(p=2\).
Let 
\begin{equation}
a_i\equiv \Gamma_i,\qquad \im b_i\equiv -\Delta_i, \qquad \mbox{ for \(p=2\).} 
\end{equation}
Then, from Eq. \eqref{uvs2pfs},
\begin{eqnarray}
a_i=\sigma^x_i \Big(\prod_{j=1}^{i-1} \sigma^z_j\Big),\qquad 
b_i=\sigma^y_i \Big(\prod_{j=1}^{i-1} \sigma^z_j\Big),
\end{eqnarray}
so that \(a_i^\dagger=a_i,\  b_i^\dagger=b_i\), 
and Eqs. \eqref{omegacommutation}, \eqref{omegacommutation2}, and 
\eqref{power} can be recast in
the form most familiar for a Clifford algebra
\begin{eqnarray}
\{a_i,a_j\}=2\delta_{i,j},\qquad 
\{b_i,b_j\}=2\delta_{i,j}, \qquad \{a_i,b_j\}=0\ .
\end{eqnarray}
Regarded as quantum degrees of freedom, the operators
\(a_i,\ b_i\) are known as {\it Majorana fermions}, or
just Majoranas for short, in recent condensed matter and quantum computation literature 
\cite{ivanov,kitaev_wire,majoranas,shusa,carlo}. 

There is one key piece of information that sets a drastic divide between
\(p=2\) and \(p\geq3\). Following the relations 
\begin{equation}\label{m2f}
a_i=C_i+C_i^\dagger\ ,\qquad \im b_i=C_i-C^\dagger_i\ , 
\end{equation}
Majoranas can be combined to yield creation \(C^\dagger_i\) 
and annihilation \(C_i\) operators of fermionic particles, 
\begin{equation}\label{cfrels}
\{C_i^{\;},C_j^\dagger\}=\delta_{i,j},\quad 
\{C_i,C_j\}=0=\{C_i^\dagger,C_j^\dagger\},\qquad i,j=1,\dots,M.
\end{equation}
It is textbook knowledge that this algebra, i.e., 
the canonical anti-commutation relations of Eq. \eqref{cfrels},  
determines completely the Fock space for indistinguishable fermions with 
\(M\) available orbitals \cite{manybody}.
So even though Majorana fermions are not particles (not even in the effective
sense of Landau quasiparticles), they are tightly linked to 
fermionic particles and the associated fermionic Fock space.
In Section \ref{Fock_parafermions} we will show that there exists a Fock
space representation of parafermions for all \(p\). The associated
indistinguishable particles satisfy unconventional exclusion and exchange
statistics. We call these particles {\it Fock parafermions}. 

Parafermions can be defined in more than one spatial dimension, provided one 
establishes an order for 
the set of indices labelling the Weyl generators.
Suppose for concreteness that Weyl generators \(V_\r, U_\r\) are placed
at each site \(\r\) of a hypercubic lattice of finite extent \(M\) in 
each of its \(d\) spatial directions, that is, \(\r=(m^1,\dots,m^d)\) with 
\(m^\mu=1,\dots,M\), $\mu=1,2,\cdots,d$. Then the sites \(\r\) can be ordered
lexicographically. For example, in $d=2$ dimensions, a possible order is
\begin{equation}
(1,1)<(1,2)<\dots<(1,M)<(2,1)<\dots<(M,M-1)<(M,M) . \nonumber
\end{equation}
Once an order is established, parafermions 
can be defined as
\begin{equation}\label{uvs2pfsn}
\Gamma_\r= V_\r \prod_{\x <\r} U_{\x}, \qquad 
\Delta_\r= V_\r U_\r  \prod_{\x<\r} U_{\x},
\end{equation}
in any number of spatial dimensions, and still satisfy the generalized Clifford
algebra of Eqs. \eqref{omegacommutation}, \eqref{omegacommutation2}
with  respect to that chosen order. For the special
case \(p=2\), these considerations together with Eq. \eqref{m2f}
lead directly to the Jordan-Wigner transformation for Majorana fermions
in more than one space dimension \cite{batista_ortiz}. 

Parafermions have a specially transparent physical interpretation
in one dimension where they arise in the study of  
self-dual clock models \cite{pclock,conII}. 
The simplest instance is the vector Potts \cite{Note2}, or $p$-clock, model 
\cite{pclock,conII}
\begin{equation}
H_{\sf VP}[h_i,J_i]=-\frac{1}{2}
\big[\sum_{i=1}^Mh_i  \, U_i +\sum_{i=1}^{M-1} J_{i} \,
V_{i}^{\;}V_{i+1}^\dagger+J_M\, V_M \big ]+{\sf h.c.}\ .
\label{pclockH}
\end{equation} 
The boundary term \(J_M\, (V_M+V_M^\dagger)\) 
enforces  boundary conditions chosen to showcase the presence of an
exact symmetry, not of the Hamiltonian but rather of the model's 
bond algebra of interactions \cite{con,conII}. Let  
\begin{equation}\label{rofk}
r(i)=M+1-i\ .
\end{equation}
denote the reflection through the system's midpoint. Then the mapping
\begin{eqnarray}
U_1 &\dual& V_{r(1)}^\dagger=V_M^\dagger\ ,\\
U_i &\dual& V_{r(i)}^\dagger V_{r(i)+1}^{\;}\ ,\qquad i=2,\dots,M,\\
V_{i}^{\;}V_{i+1}^\dagger &\dual& U_{r(i)}\,  \qquad 
\quad \qquad \ i=1,\dots,M-1, \\
V_M &\dual& U_{r(M)}=U_1\ ,
\end{eqnarray} 
local in the model's ``interactions" (or more precisely, bonds \cite{conII}), 
induces an algebra isomorphism \(\Phi_\d\). According to the general
theory of dualities developed in Ref. \cite{conII}, the mapping
\(\Phi_\d\) is {\it unitarily implementable}. This means that there exists
a unitary transformation \(\mathcal{U}_\d\) such that 
\begin{equation}
\mathcal{O}\dual \widehat{\mathcal{O}},\qquad 
\widehat{\mathcal{O}}\equiv\Phi_\d(\mathcal{O})=
\mathcal{U}_\d \mathcal{O}\mathcal{U}_\d^\dagger
\end{equation}
for any operator \(\mathcal{O}\). 

The importance of these observations follow from the effect of the duality
mapping on the vector Potts hamiltonian. Since
\begin{equation}
H_{\sf VP}[h_i,J_i] \dual H_{\sf VP}[h^*_i,J^*_i]\ ,
\end{equation}
with dual couplings
\begin{equation}
h_i^*\equiv J_{r(i)},\qquad J_i^*\equiv h_{r(i)} \ ,
\end{equation}
it follows that 
\begin{equation}
H_{\sf VP}[h^*_i, J^*_i]=\mathcal{U}_\d \, H_{\sf VP}[h_i,J_i] \, 
\mathcal{U}_\d^\dagger\ .
\end{equation}
That is , any pair of vector Potts Hamiltonians associated to
points \(\{h_i,J_i\}\) and \(\{h_i^*,J_i^*\}\) in coupling
space are isospectral. In more physical terms, the  vector Potts model 
is {\it self-dual}. 

Now that we understand the model's self-duality in detail 
we can restore conventional open boundary conditions by setting
\(J_M=0\). By restoring conventional boundary conditions we also recover
the the model's global \(\mathds{Z}_p\) symmetry,
\begin{equation}\label{pcharge}
Q_p=\prod_{i=1}^M U_i\ ,\qquad [H_{\sf VP}[J_M=0],Q_p]=0. 
\end{equation}
The dual Hamiltonian however 
\begin{equation}
H_{\sf VP}[h_1^*=0]=\mathcal{U}_\d \, H_{\sf VP}[J_M=0] \, \mathcal{U}_\d^\dagger
\end{equation}
is not globally symmetric, but rather has a boundary symmetry as
can be checked explicitly. From the point of view of the duality transformation this 
follows because
\begin{equation}
Q_p\dual V_1^\dagger\ ,
\end{equation}
and so \(V_1^\dagger\) must commute with \(H_{\sf VP}[h_1^*=0]\).
This localized symmetry is an elementary example of a {\it holographic symmetry} 
\cite{cobanera2013}, that is, a boundary symmetry dual to a bulk symmetry.
As explained in Ref. \cite{cobanera2013}, holographic symmetries are valuable 
stepping stones in the search for interacting zero-energy edge modes and
generalized order parameters for topologically-ordered quantum phases 
of matter.

Since the duality mapping \(\Phi_\d\) is an isomorphism, the dual variables
\begin{eqnarray}
\widehat{V}_i\equiv\Phi_\d(V_i),\qquad \widehat{U}_i\equiv  \Phi_\d(U_i)
\end{eqnarray}
satisfy all the relations expected of a set of Weyl generators. Moreover,
in view of this paper's notation for the vector Potss model \cite{Note2},
the quantum variable \(V_i\) is associated to the model's order parameter.
Hence it is conventional to call the \(V_i\) order variables,
and the 
\begin{eqnarray}
\widehat{V}_i = \Phi_\d((V_i^{\;}V_{i+1}^\dagger)\dots (V_{M-1}^{\;}V_M^\dagger) V_M^{\;})
= U_{r(i)}\dots U_1   
\end{eqnarray}
{\it disorder variables} (see \cite{conII} for the relation of disorder variables
to disorder parameters). Recalling the definition of parafermions, Eq. \eqref{uvs2pfs}, 
we see that (\(\widehat{V}_{M+1}\equiv \mathds{1}\))
\begin{eqnarray}
\Gamma_i=V_i\widehat{V}_{r(i)+1},\quad \Delta_i=V_i\widehat{V}_{r(i)}, \qquad i=1,\dots,M.
\end{eqnarray}
That is, in one-dimension, 
{\it parafermions arise as the product of an order with a disorder variable}. 
This remarkable connection between self-duality and parafermions 
was first noticed in the early days of the operator product expansion,
in the context of {\it classical} statistical mechanics \cite{kadanoff1971,fradkin1980},
and later extended to quantum clock models in Ref. \cite{conII}. 

In closing, let us notice that
the \(\mathds{Z}_p\)-symmetric (\(J_M=0\)) vector Potts model can be
rewritten as a local (quadratic) Hamiltonian of parafermions,    
\begin{equation}\label{parafermionic_vp}
H_{\sf VP}=-\frac{1}{2}
\big[\sum_{i=1}^M h_i \, \Gamma_i^{\dagger}\Delta_i^{\;}+\sum_{i=1}^{M-1} J_i \,
 \Delta_{i}^{\;}\Gamma_{i+1}^\dagger  \big]+{\sf h.c.}\ .
\end{equation}

\section{Weyl algebras, parafermions, and 
self-dual representations of the braid group}
\label{sdreps_section}

Majoranas may emerge as effective excitations or
localized zero-energy modes in some topologi\-cally-ordered 
electronic phases of matter, and several current experiments 
on nanowires are specifically designed to hunt  for them 
\cite{churchill2013}.
Starting with the key observation 
that Majoranas realize a two-dimensional representation of the braid 
group \cite{ivanov},
protocols have been developed to braid Majoranas 
and experimentally achieve, with some degree of topological protection, 
(non-universal) Clifford gates 
\cite{kitaev_wire,dassarma2008,hyart2013}. 

Since Majorana fermions are but the special $p=2$ instance of parafer\-mions, 
it is natural to search for platforms and protocols that exploit 
parafer\-mions with \(p>2\) for fault-tolerant quantum information 
processing \cite{pachos,q_compu,clarke,lindner}. A necessary first step 
in this direction is to find parafermionic 
representations of the braid group. As a general rule,
these representations should be as physically motivated as possible
to foster experimental realization. Two conditions come immediately
to mind: The representations should be {\it unitary} and {\it local} in terms of
parafermions. These conditions are motivated by  
general features of the quantum control model of quantum computation \cite{q_compu}. 
We will add to this list of generic conditions  
a third one specific to parafermions: {\it the representations should be 
self-dual}.  

As explained in  previous sections, the link between parafermions, order, and
dual disorder variables offers one of the most compelling physical 
realizations of parafermions. In the following 
we will outline a systematic search 
of parafermionic representations of the braid group, for all values of \(p\), that transform
in a natural way, Eq. \eqref{dual_rep}, under duality. These are the 
{\it self-dual representations
of the braid group} alluded to. 

In this paper, by {\it Artin braid group}, or simply  {\it braid group}, 
we mean the group \(B_L\) with $L-1$ generators $\sigma_k$ and standard 
``braid relations''  \cite{turaev}
\begin{equation}
B_L=\langle \sigma_1,\dots,\sigma_{L-1}\ |\  
\sigma_k\sigma_{k+1}\sigma_k=
\sigma_{k+1}\sigma_k\sigma_{k+1},\  
\sigma_k\sigma_l=\sigma_l\sigma_k\ \mbox{if}
\ |k-l|\geq 2\rangle \ ,
\end{equation}
associated to the quantum-mechanical description of \(L\) indistinguishable
any\-ons moving in the infinite plane \(\mathds{R}^2\). The 
mapping 
\begin{equation}\label{braiding_auto}
\sigma_{k}\mapsto \sigma_{L-k},\qquad k=1,\dots,L-1 ,
\end{equation}
preserves all the braiding relations and so 
induces an automorphism of \(B_L\) that we will have the occasion
to use later in this section. 

Next we search for unitary representations of the braid group 
\(B_{2M+1}\) of the form 
\begin{eqnarray}
\rho^{(p)}_{\sf sd}(\sigma_{2i-1})&=&
\frac{1}{\sqrt{p}}\sum_{m=0}^{p-1} \alpha_{m} U_i^{\dagger m}\ ,\qquad
\qquad \ i=1,\dots,M, \\
\rho^{(p)}_{\sf sd}(\sigma_{2i})&=& 
\frac{1}{\sqrt{p}}\sum_{m=0}^{p-1} \beta_{m} (V^{\;}_iV_{i+1}^{\dagger})^m\ ,
\qquad i=1,\dots,M-1,\\
\rho^{(p)}_{\sf sd}(\sigma_{2M})&=& 
\frac{1}{\sqrt{p}}\sum_{m=0}^{p-1} \beta_{m} (V^{\;}_{M})^m\ .
\end{eqnarray}
The complex parameters \(\alpha_{m}, \beta_{m},\ m=0,\dots,p-1,\) are constrained
by two requirements: unitarity,
\begin{equation}
\big(\rho^{(p)}_{\sf sd}(\sigma_{k})\big)^\dagger=\rho^{(p)}_{\sf sd}(\sigma_{k}^{-1})\ ,
\end{equation}
and the braiding relations
\begin{equation}
\rho^{(p)}_{\sf sd}(\sigma_k)\rho^{(p)}_{\sf sd}(\sigma_{k+1})
\rho^{(p)}_{\sf sd}(\sigma_{k})=
\rho^{(p)}_{\sf sd}(\sigma_{k+1})\rho^{(p)}_{\sf sd}(\sigma_{k})
\rho^{(p)}_{\sf sd}(\sigma_{k+1})\ .
\end{equation}
Unitarity implies the set of equations
(the overbar denotes complex conjugation)
\begin{equation}\label{unitarity_poly}
\sum_{m=0}^{p-1}\alpha_{q+m}\bar{\alpha}_{m}=
\sum_{m=0}^{p-1}\beta_{q+m}\bar{\beta}_{m}=\left \{
\begin{array}{lc}
p &\mbox{if}\ \ q=0\\
0 & \mbox{\ otherwise}
\end{array} \right. \ .
\end{equation}
The braiding relations on the other hand require that
\begin{equation}\label{braiding_rel}
\alpha_r\sum_{m+n=q\ ({\sf mod}\ p)}\omega^{rm}\beta_m\beta_n=
\beta_q\sum_{m+n=r\ ({\sf mod}\ p)}\omega^{qm}\alpha_m\alpha_n,
\quad r,q=0,\dots,p-1.
\end{equation}

We call (the representation associated to) any non-trivial solution of
Eqs. \eqref{unitarity_poly} and \eqref{braiding_rel} a self-dual 
representation of the braid group. Self-dual representations 
are physically distinguished because they transform in a natural fashion under 
the duality mapping \(\mathcal{U}_\d\) described
in Section \ref{Weyl_parafermions}. By this we mean the following.
The duality transformation \(\mathcal{U}_\d\) can be combined with 
the automorphism of the braid group Eq. \eqref{braiding_auto} to define
out of any given self-dual representation a new representation \(\rho^D\),
\begin{equation}\label{dual_rep}
\rho^D(\sigma_k)\equiv \mathcal{U}_\d\, \rho^{(p)}_{\sf sd}(\sigma_{r(k)+M})\,
\mathcal{U}_\d^\dagger 
\end{equation}
(\(r(k)\) was defined in Eq. \eqref{rofk}).
The point we want to emphasize is that \(\rho^D\) is again of the self-dual form,
\begin{eqnarray}
\rho^D(\sigma_{2i-1})&=&
\frac{1}{\sqrt{p}}\sum_{m=0}^{p-1} \alpha^*_m U_i^{\dagger m}\ ,\qquad
\qquad \ i=1,\dots,M, \\
\rho^D(\sigma_{2i})&=& 
\frac{1}{\sqrt{p}}\sum_{m=0}^{p-1}  \beta^*_m  (V^{\;}_iV_{i+1}^{\dagger})^m\ ,
\qquad i=1,\dots,M-1,\\
\rho^D(\sigma_{2M})&=& 
\frac{1}{\sqrt{p}}\sum_{m=0}^{p-1}  \beta^*_m  (V^{\;}_{M})^m\ ,
\end{eqnarray}
 with dual ``couplings"
\begin{equation}
\alpha^*_m=\beta_{m},\quad \beta^*_m=\alpha_{p-m},\qquad m=0,\dots,p-1.
\end{equation}

A self-dual representation of \(B_{2M}\) can be obtained
by dropping \(\rho^{(p)}_{\sf sd}(\sigma_{2M})\). 
From the point of view of measurement-assisted topological quantum computation,
these representations are the most interesting ones because they are local
and quadratic in terms of parafermions,
\begin{eqnarray}
\rho^{(p)}_{\sf sd}(\sigma_{2i-1})=
\frac{1}{\sqrt{p}}\sum_{m=0}^{p-1} \alpha_{m} (\Delta_i^\dagger\Gamma_i^{\,})^m\ ,\quad
\rho^{(p)}_{\sf sd}(\sigma_{2i})= 
\frac{1}{\sqrt{p}}\sum_{m=0}^{p-1} \beta_{m} (\Delta_i^{\;}\Gamma_{i+1}^\dagger)^m\ .
\label{sd_braiding_parafermions}
\end{eqnarray}
They are also  reducible, since they commutes with the
\(\mathds{Z}_p\) charge operator \(Q_p\) of Eq. \eqref{pcharge}. It is useful
to notice that the projectors \(P_q=P_q^2=P_q^\dagger\) onto the sectors of total
\(\mathds{Z}_p\) charge \(\omega^q\) can be computed as
\begin{equation}
P_q=\frac{1}{p}\sum_{m=0}^{p-1}\bar{\omega}^{mq}Q_p^m\ .
\end{equation}

Self-dual representations have two very useful general properties. 
First, recall the relative prime decomposition of \(p\), Eq. \eqref{p_decomp},
and suppose we have self-dual representations \(\rho^{(r_j)}_{\sf sd}\) with
coefficients \(\alpha^{(r_j)}_m,\beta^{(r_j)}_m,\ m=0,\dots,r_j-1\). 
Then one can exploit the canonical embedding, Eq. \eqref{canonical_embedding}, 
to transform the self-dual representations \(\rho^{(r_j)}_{\sf sd}\) of dimensions
\(r_j\) to self-dual representations of dimension \(p\),
\begin{equation}
\rho^{(r_j)}_{\sf sd}\hookrightarrow \rho^{(p)}_{{\sf sd},j}\ ,
\end{equation}
with 
\begin{eqnarray}\label{embedded_reps}
\rho^{(p)}_{{\sf sd},j}(\sigma_{2i-1})&=&
\frac{1}{\sqrt{r_j}}\sum_{m=0}^{r_j-1} 
\alpha^{(r_j)}_{m} (U_i^{\dagger})^{ mp/r_j}\ ,\qquad
\qquad i=1,\dots,M, \\
\rho^{(p)}_{{\sf sd},j}(\sigma_{2i})&=& 
\frac{1}{\sqrt{r_j}}\sum_{m=0}^{r_j-1} \beta^{(r_j)}_{m} 
(V^{\;}_iV_{i+1}^{\dagger})^{mp^2/r_j^2}\ ,
\qquad i=1,\dots,M-1.
\end{eqnarray}
A key characteristic of these embedded representations is the 
{\it absence} of certain powers of the Weyl generators in the expansion of 
\(\rho^{(p)}_{{\sf sd},j}\), that is, the vanishing of some of the coefficients
\(\alpha_m\) and/or \(\beta_m\).

Second, the tensor product of two self-dual representations
of dimensions \(p_1\), \(p_2\) is canonically isomorphic to a self-dual representation
of dimension \(p_1p_2\), {\it provided \(p_1,p_2\) are relative primes}. We can
state this property most elegantly in terms of the mapping \(\cong\) of 
Eq. \eqref{cong_otimes1},
\begin{equation}\label{compose_to_sd}
\rho^{(p_1)}_{\sf sd}\otimes\rho^{(p_2)}_{\sf sd}\cong\rho^{(p_1p_2)}_{\sf sd}\ .
\end{equation}
This canonical identification preserves a very important property. Suppose
neither \(\rho^{(p_1)}_{\sf sd}\) nor \(\rho^{(p_2)}_{\sf sd}\) contain 
vanishing coefficients. Then \(\rho^{(p_1p_2)}_{\sf sd}\)
on the right-hand side of Eq. \eqref{compose_to_sd} contains no vanishing coefficients
either. 

The next crucial task is to actually find self-dual representations, that is,
solutions of Eqs. \eqref{unitarity_poly} and \eqref{braiding_rel}. We will
focus on solutions without vanishing coefficients, since we know that solutions
with vanishing coefficients have an interpretation in terms of embedded 
representations. 

\subsection{p odd}

For \(p=3,5,\dots\) odd it is convenient to impose the extra constraint  
\begin{equation}\label{aequalsb}
\alpha_m=\beta_m,\qquad m=0,\dots,p-1.
\end{equation}
Then there exist \(2p\) solutions with no vanishing coefficients. 
However, it suffices to consider only the conjugate pair of solutions
\begin{equation}\label{podd_sols}
\alpha_{m}=\beta_m=\left\{
\begin{array}{l}
\omega^{m(m-1)/2} \\
\bar{\omega}^{m(m-1)/2}
\end{array} \right.  \ .
\end{equation}
Any other one of the remaining \(2(p-1)\) 
determines a self-dual representation equivalent to one of these two.
Furthermore, if \(p\) is a prime of the form \(p=4n+1\), the two solutions of
Eq. \eqref{podd_sols} determine unitarily equivalent representations.

The special role of \(p=4n+1\) prime can be understood as follows.
The sequence of odd primes can be split into two subsequences 
of the form \(4n+1\) (\(5,13,17,\dots\)) and \(4n+3\) (\(3,7,11,\dots\)).
Also, for \(p\) prime (and only for \(p\) prime), the set of remainders modulo $p$, 
\(\mathds{Z}_p={0,1,\dots,p-1}\), becomes a discrete field under modular addition and 
multiplication. But the polynomial equation \(x^2+1\equiv 0\) (modulo \(p\)) has a 
solution in \(\mathds{Z}_p\) only if \(p\) belongs to the \(4n+1\) sequence. 

Note  that the Gaussian
representation of the braid group described in Ref. \cite{j} can
be reinterpreted as a self-dual representation with 
\(\alpha_m\neq\beta_m\). The Gaussian representation \cite{j,gj} is defined
for any \(p\) odd as
\begin{equation}
\rho_{\sf G}^{(p)}(\sigma_k)=\frac{1}{\sqrt{p}}
\sum_{m=0}^{p-1}\omega^{m^2} u_k^m\ ,\qquad k=1,\dots,L-1,
\end{equation} 
(we change the normalization relative to Ref. \cite{j})
in terms of \(L-1\) generators $u_k$ satisfying
\begin{eqnarray}
u_k^{p}&=&\mathds{1},\qquad \qquad \ \ \ \ \, k=1,\dots,L-1,\\
u_ku_{k+1}&=&\omega^2 \, u_{k+1}u_k,\qquad k=1,\dots,L-2,
\end{eqnarray}
and commuting otherwise. There exists a faithful, irreducible representation
of these relations in terms of Weyl generators,
\begin{equation}
u_{k}\mapsto \left\{
\begin{array}{lcl}
U_i^{\dagger 2} & \mbox{if} & k=2i-1,\ \ i=1,\dots,M\\
V^{\;}_iV_{i+1}^{\dagger}& \mbox{if} & k=2i,\ \ \ \ \ \ \ i=1,\dots,M-1\\
V_{M}& \mbox{if} & k=2M
\end{array} \right. 
\end{equation}
with \(L=2M+1\) as before. Then the Weyl realization of the Gaussian 
representation is self-dual, 
\begin{eqnarray}
\rho^{(p)}_{\sf G}(\sigma_{2i-1})&=&
\frac{1}{\sqrt{p}}\sum_{m=0}^{p-1} \alpha_{m} U_i^{\dagger m}\ ,\qquad
\qquad \ i=1,\dots,M, \\
\rho^{(p)}_{\sf G}(\sigma_{2i})&=& 
\frac{1}{\sqrt{p}}\sum_{m=0}^{p-1} \beta_{m} (V^{\;}_iV_{i+1}^{\dagger})^m\ ,
\qquad i=1,\dots,M-1,\\
\rho^{(p)}_{\sf G}(\sigma_{2M})&=& 
\frac{1}{\sqrt{p}}\sum_{m=0}^{p-1} \beta_{m} (V^{\;}_{M})^m\ ,
\end{eqnarray}
with 
\begin{eqnarray}
\beta_m=\omega^{m^2}=\alpha_{2m\ ({\sf mod}\ p)},\qquad m=0,\dots,p-1.
\end{eqnarray}
Unlike our previous solution, Eq. \eqref{podd_sols}, the Gaussian solution has
\(\alpha_m\neq\beta_m\), but the two sets of coefficients are connected
by a permutation. The reason is that, since \(p\) is assumed to be odd, 
\(2\) and \(p\) are relative primes. Then, the mapping 
\begin{equation}
m \mapsto 2m \ ({\sf mod}\ p),\qquad m=0,\dots,p-1,
\end{equation}
is one-to-one and onto, that is, a permutation.

The fact that the Gaussian representation is self-dual could potentially
have practical implications, since there exist recent proposals 
to realize the Gaussian representation of \(B_{2M}\) in mesoscopic 
arrays  \cite{lindner,hastings}.

\subsection{\(p=2^n\), \(n=1,2,\dots\)}

While we have reasons to believe that self-dual representations with 
\(p=2^n\) follow a pattern analogous to the one found for \(p\) odd,
it remains an open problem to find the non-trivial solutions of 
Eqs. \eqref{unitarity_poly} and \eqref{braiding_rel} for arbitrary \(p=2^n\). 
For small \(n\) it is possible to solve the equations by elementary algebra.
Again we impose the extra constraint Eq. \eqref{aequalsb}.

For \(n=1\), that is, \(p=2\), we find only two solutions
\begin{equation}\label{majorana_table}
\begin{array}{c}
 \begin{tabular}{ | c | c |}
    \hline
    \(\alpha_0\) & \(\alpha_1\) \\ \hline \hline 
    1 & \(\im\) \\ \hline
    1 & -\(\im\) \\ 
    \hline
  \end{tabular}
\end{array}
\end{equation}
They define unitarily equivalent self-dual representations. Recalling
the connection between Majorana fermions and parafermions for \(p=2\), we obtain that
our self-dual representation reduces to the Majorana representation
\begin{equation}\label{majorana_reps}
\rho^{(2)}_{\sf sd}(\sigma_{2i-1})=\frac{\mathds{1}\mp b_ia_i}{\sqrt{2}},\qquad
\rho^{(2)}_{\sf sd}(\sigma_{2i})=\frac{\mathds{1} \mp a_{i+1}b_i}{\sqrt{2}}
\end{equation}
first described in Ref. \cite{ivanov}. (The two signs correspond to the two 
unitarily equivalent representations of the table above.) This representation of the 
braid group amounts to a {\it Fock representation}, since it can be rewritten 
in terms of creation and annihilation operators for ordinary fermions,
\begin{eqnarray}
\rho^{(2)}_{\sf sd}(\sigma_{2i-1})&=&\frac{\mathds{1}\mp\im 
(2C^\dagger_iC_i^{\;}-\mathds{1})}{\sqrt{2}}\ ,\nonumber \\
\rho^{(2)}_{\sf sd}(\sigma_{2i})&=&
\frac{\mathds{1}\pm\im (C_i^\dagger C_{i+1}^{\;}+C_{i+1}^\dagger
C_i^{\;}+ C_{i+1}C_i+C_{i}^\dagger C_{i+1}^\dagger)}{\sqrt{2}}\ . 
\label{braiding_fermions}
\end{eqnarray}
Notice the emergence of the anomalous pairing term 
\(C_iC_{i+1}+C_{i}^\dagger C_{i+1}^\dagger\).

Let us illustrate in passing the notion of embedded representation. 
Suppose for example that \(p=2q\), with \(q\) odd. Then from 
Eqs. \eqref{embedded_reps} and \eqref{majorana_table} we obtain
the \(p\) dimensional representations
\begin{eqnarray}
\rho^{(2q)}_{\sf sd}(\sigma_{2i-1})&=&\,
\ \ \ \frac{\mathds{1}\pm\im U_i^{\dagger q }}{\sqrt{2}}\ \ \ \, =
\frac{\mathds{1}\pm\im (\Delta_i^\dagger\Gamma_i^{\;})^q }{\sqrt{2}}\ ,\\
\rho^{(2q)}_{\sf sd}(\sigma_{2i})&=&
\frac{\mathds{1}\pm\im (V_iV_{i+1}^\dagger)^q}{\sqrt{2}}=
\frac{\mathds{1}\pm\im (\Delta_i^{\;}\Gamma_{i+1}^{\dagger})^q}{\sqrt{2}}\ ,
\end{eqnarray}
with only two non-vanishing coefficients.

For \(n=2\), that is, \(p=4\), we find two classes of solutions,
each class containing four possible sets of coefficients,
\begin{equation}
\begin{array} {lr}
\begin{tabular}{| c | c | c | c |}
    \hline
    \(\alpha_0\) & \(\alpha_1\) & \(\alpha_2\) & \(\alpha_3\) \\ \hline\hline
    1 & \(\im\)  & -1   & \(\im\)  \\ \hline
    1 & -\(\im\) & -1   & -\(\im\) \\ \hline
    1 &  -1       & 1   & 1 \\\hline
    1 &   1       & 1   & -1 \\ 
    \hline
  \end{tabular} 
\qquad \qquad & \qquad \qquad
\begin{tabular}{ | c | c | c | c | }
    \hline
    \(\alpha_0\) & \(\alpha_1\) & \(\alpha_2\) & \(\alpha_3\) \\ \hline\hline
    1 & \(\im\)  &  1   & -\(\im\) \\  \hline
    1 & -\(\im\) &  1   & \(\im\) \\ \hline
    1 &     1      &  -1   & 1 \\ \hline
    1 &     -1      &  -1   & -1 \\ 
    \hline
  \end{tabular}
\end{array}
\end{equation}
The representations in a given class are unitarily equivalent. Moreover, 
the representation associated to the left class of solutions are unitarily
equivalent the product representation \(\rho^{(2)}_{\sf sd}\otimes 
\rho^{(2)}_{\sf sd}\).

The conclusion of the previous paragraph suggests an interesting 
conjecture. It may be that for arbitrary \(n\), \(p=2^n\), there exist
\(n\) classes of solutions of Eqs. \eqref{unitarity_poly} and \eqref{braiding_rel}, 
with one class at least unitarily equivalent to the product representation
\begin{equation}\label{product_of_2s}
\rho^{(2)}_{\sf sd}\otimes \dots \otimes \rho^{(2)}_{\sf sd}\qquad (n\ \mbox{factors}).
\end{equation}
If this is true, it poses a potentially non-trivial constraint on that class of 
representations \(\rho_{\sf sd}^{(2^n)}\). The reason is that the 
representation   \(\rho_{\sf sd}^{(2)}\) has finite period \(8\), 
\begin{equation}
\big(\rho^{(2)}_{\sf sd}(\sigma_k)\big)^8=\rho^{(2)}_{\sf sd}(\sigma_k^8)=\mathds{1},
\qquad \forall k\ .
\end{equation}
It follows that the product representation \eqref{product_of_2s} has 
period  \(4\) for \(n\) even, or period \(8\) for \(n\) odd.
If one exists, the same must hold for an equivalent self-dual 
representation \(\rho_{\sf sd}^{(2^n)}\).

\subsection{General \(p\)}

It will be easy to construct interesting solutions for arbitrary \(p\)
once the special case \(p=2^n\) is completely solved. The reason 
is that an an arbitrary \(p\) can always be decomposed as \(p=2^n q\), with 
\(q\) odd. This is fortunate, because \(2^n\) and \(q\) are by definition
relative primes, and so it is possible to 
use Eq. \eqref{compose_to_sd} to obtain a self-dual representation 
(with no vanishing coefficients!) for any p,
\begin{equation} 
\rho_{\sf sd}^{(2^n)}\otimes \rho_{\sf sd}^{(q)}\cong \rho_{\sf sd}^{(2^nq)}\ .
\end{equation}
The resulting representation on the right-hand side will
not in general satisfy the extra condition Eq. \eqref{aequalsb},
even if \(\rho^{(2^n)}_{\sf sd}\) and \(\ \rho^{(q)}_{\sf sd}\) do satisfy it
individually. Specially interesting are the representations obtained 
for \(p=4q\) combining our solutions by this procedure.

\section{Fock parafermions}
\label{Fock_parafermions}

The Majorana representation of the braid group Eq. \eqref{majorana_reps}
is one of the most studied in physics \cite{ivanov,carlo}. We think this is in
part because it has a Fock space interpretation in terms of ordinary fermions,
Eq. \eqref{braiding_fermions}. The key point is that this interpretation immediately 
suggests the right quantum state of matter to naturally support this type of braiding: 
electron vacua with anomalous pairing. These
observations are typically summarized in the literature as conditions for
the emergence of Majorana fermions in condensed matter systems \cite{kitaev_wire,carlo}. 

Parafermions with \(p\geq3\) offer a natural generalization of the Majorana 
braiding paradigm. As we established in the previous Section, 
there exist a variety of representations 
of the braid group naturally linked to parafermions through 
Eq. \eqref{sd_braiding_parafermions}. There also exist proposals to realize parafermions as 
zero-energy modes \cite{fendley} in mesoscopic arrays \cite{clarke,lindner,bvhc}. 
Hence the following question becomes relevant: Do particles 
with exotic $p$-exclusion, $\theta=2\pi/p$-exchange statistics,
{\it and} a well defined associated Fock space exist?
In this section we will answer this question in the affirmative by introducing  Fock spaces 
\(\mathcal{F}_p(M)\) of indistinguishable particles satisfying the required
statistical conditions (\(M\) is the number of available orbitals). We call
these particles {\it Fock parafermions}. For \(p=2\), Fock parafermions
are just ordinary fermions, but for \(p>2\) they are anyons with 
unconventional exclusion and exchange statistics. 
One of our goals is to obtain an expansion of parafermions
in terms of creation and annihilation operators of Fock parafermions.
 
\subsection{The Fock space of Fock parafermions}

In this paper we propose to construct
the Fock space of of indistinguishable, {\it independent} particles in terms of 
two fundamental pieces of information: 
\begin{enumerate}
\item 
the state space of a single particle \(\mathcal{H}\) (the available orbitals), and
\item
a rule to multiply \(N\) single-particle states to generate an \(N\)-body state
{\it with the correct exchange and exclusion statistics.}
\end{enumerate}
Hence a Fock space is a state space (a Hilbert space defined over the field 
of the complex numbers $\mathbb{C}$) endowed with
a physically motivated, {\it associative} multiplication. Once a Fock space is
specified in this fashion, the unintuitive in general algebra of creation and 
annihilation operators can be {\it derived} systematically. In particular, we will see that
the algebra of creation and annihilation operators for Fock
parafermions displays a variety of unconventional features that would have
been extremely hard to guess {\it a priori}. 

Let us start by specifying the single-particle state space \(\mathcal{H}\) 
by choosing a basis of orthonormal orbitals \(\phi_1,\dots,\phi_M\), and
suppose for now that the Fock parafermions are {\it independent} 
particles (we can always add interactions later). Then we can specify a 
many-body state by stating that there are \(n_1\) Fock parafermions 
occuppying orbital \(\phi_1\), \(n_2\) Fock parafermions 
occuppying orbital \(\phi_2\) and so on. Mathematically, we can organize
the occupation numbers into a unique object, the {\it ordered} list
\((n_1,\dots,n_M)\), and assign to this list a unique (up to a phase that 
we ignore in the following) many-body state,
\begin{equation}\label{assignment}
(n_1,\dots,n_M)\mapsto |n_1,\dots,n_M\rangle\qquad \ . 
\end{equation} 
Then the general structure of the Fock space is 
\begin{equation}
\mathcal{F}_p(M)=\bigoplus_{N=0}\ {\sf Linear\ Span}
\{|n_1,\dots,n_M\rangle|\sum_{r=1}^Mn_r=N\},
\end{equation}
with inner product  
\begin{equation}\label{mb_inner}
\langle {n_1}, \cdots, {n_M}|{n_1'},\cdots,{n_M'}\rangle
=\prod_{i=1}^M\delta(n_i,n_i')\ .
\end{equation}
The inner product is strongly physically motivated. It
guarantees that states with distinct orbital occupations are orthogonal. 

This minimal description of 
the Fock space of Fock parafermions contains no statistical
information. First, the range
\[
n_i=0, 1, \dots,n_{\sf E}, \ \ \  i=1,\cdots,M ,
\]
of the occupation numbers cannot be specified until we determine
the exclusion statistics of Fock parafermions.
We call the not necessarily finite integer \(n_{\sf E}\geq 2\) the exclusion parameter.
Second, the state of a system of independent Fock parafermions is 
{\it uniquely specified} (up to a phase) by the mapping Eq. \eqref{assignment}.
In this formalism, the notion of exchanging two Fock parafermions has no
mathematical representation.
Thus the present description of \(\mathcal{F}_p(M)\) contains no
information on the exchange statistics of Fock parafermions.
In the following we will use the abbreviated notation
\begin{eqnarray}
|n_i\rangle&=&|0,\dots,0,n_i,0,\dots,0\rangle\ ,\nonumber\\ 
|n_i,n_j\rangle&=& |0,\dots,0,n_i,0,\dots,0,n_j,0,\dots,0\rangle\ ,\nonumber
\end{eqnarray}
and so on.

To account for both aspects of particle statistics, we 
define an {\it associative} multiplication \(\times\) in Fock space
in two steps. First, the rule
\begin{equation}\label{powers}
(|n_i=1\rangle)^m\equiv|n_i=m\rangle=|n_i=1\rangle\times \dots \times |n_i=1\rangle
\qquad(m\ \mbox{times})
\end{equation} 
allow us to describe algebraically the process of adding Fock parafermions
to any fixed orbital. In light of this definition,
the mathematical role of the exclusion parameter
becomes clear: the exclusion parameter \(n_{\sf E}\geq 2\) is the smallest
integer such that
\begin{equation}
(|n_i=1\rangle)^{n_{\sf E}}=|n_i=n_{\sf E}\rangle=0\ .
\end{equation}
Second, the requirement of consistency with the exchange rules for parafermions, 
Eqs. \eqref{omegacommutation} and \eqref{omegacommutation2},
suggests the definition
\begin{eqnarray}\label{2exchange} 
|n_i\rangle \times 
|n_j\rangle=\omega^{n_in_j}|n_j\rangle \times 
|n_i\rangle \equiv|n_i,n_j\rangle\qquad \mbox{for}\ \ i<j\label{exchange_rule}
\end{eqnarray} 
( $\omega=e^{\im \theta}$, $\theta=2\pi/p$) for combining two different
orbitals with arbitrary occupation. 
This multiplication rule captures the {\it exchange} statistics 
of Fock parafermions. 

With these definitions, we see that any state in Fock space can be 
generated by multiplying single-particle states as follows
\begin{equation}
|n_1,\dots,n_M\rangle=(|1,0,\dots,0\rangle)^{n_1}\times \dots\times
(|0,\dots,0,1\rangle)^{n_M}.
\end{equation}
Hence we see that the Fock vacuum 
\begin{equation}\label{omega}
|0\rangle\equiv|n_1=0,\dots,n_M=0\rangle=(|1,0,\dots,0\rangle)^{0}\times \dots\times
(|0,\dots,0,1\rangle)^{0}
\end{equation}
plays the distinguished role of multiplicative identity,
\begin{equation}\label{fock_identity}
|n_1,\dots,n_M\rangle\times|0\rangle=|0\rangle\times|n_1,\dots,n_M\rangle=
|n_1,\dots,n_M\rangle\ .
\end{equation}

Finally, we need to specify the exclusion parameter 
\(n_{\sf E}\) for Fock parafer\-mions. Remarkably, 
\(n_{\sf E}\) is determined by the exchange angle 
\(\theta\), the basic physical considerations leading to the inner 
product Eq. \eqref{mb_inner}, and the following assumption:
{\it 
any state \(|\Psi\rangle\) in Fock space that commutes with
every other state \(|\Phi\rangle \in \mathcal{F}_p(M)\) is a scalar 
multiple of the Fock vacuum,}
\begin{equation}\label{assumption}
|\Psi\rangle\times |\Phi\rangle=|\Phi\rangle\times| \Psi\rangle,
\quad \forall |\Phi\rangle\qquad \Rightarrow \qquad 
|\Psi\rangle=\alpha|0\rangle\ , \quad \alpha \in \mathds{C}\ .
\end{equation}
We can understand this assumption as the algebraic statement
reflecting the uniqueness of the Fock vacuum. 
Later we will see
that this assumption implies any operator in Fock space
that commutes with every parafermionic creation (or annihilation) operator 
is a scalar multiple of the identity operator. 

The determination of \(n_{\sf E}\) starts with the observation that
\begin{eqnarray}
|n_i=p\rangle \times |n_j\rangle=|n_j\rangle \times |n_i=p\rangle ,
\end{eqnarray}
thanks to Eqs. \eqref{powers}, \eqref{2exchange}, and \(\omega^p=1\),
and so the \(M\) states \( |n_i=p\rangle\) satisfy
\begin{equation}
|n_i=p\rangle=\alpha_i|0\rangle , \ \ \ i=1,\cdots, M ,
\end{equation}
according to our assumption, Eq. \eqref{assumption}. 
But since these states contain \(p\) particles,
\begin{equation}
0=\langle  0|n_i=p\rangle=\alpha_i\ .
\end{equation}
It follows that
\begin{equation}
|n_i=p\rangle=(|n_i=1\rangle)^p=0\ .
\end{equation} 
We conclude that \(n_{\sf E}= p\), i.e., Fock parafermions satisfy
\(p\)-exclusion and $2\pi/p$-exclusion. Thus, the dimension of the Fock space 
of Fock parafermions is
\begin{eqnarray}
\dim\ \mathcal{F}_p(M)=p^M\ .
\end{eqnarray}

In the mathematical literature, the algebra \(\mathcal{F}_p(M)\)
is known as the \(p\)-Grassmann algebra with \(M\) generators \cite{traubenberg,kwasniewski}.
Its many-body interpretation seems to have gone unnoticed up to now, for reasons
discussed in Section \ref{cs_to_gs}.
Let us notice in closing that \(\mathcal{F}_2(M)\) is just the standard Fock space of 
indistinguishable fermions with \(M\) available orbitals \cite{manybody,ballentine}. 
This will become self-evident in the next section when we compute the algebra of creation and
annihilation operators associated to \(\mathcal{F}_p(M)\). 

\subsection{Creation and annihilation operators}

By definition, a single application of a creation operator for 
orbital \(i\) adds a particle in that orbital to any many-body state.
Our proposal is to define creation operators \(C_i^\dagger\) of Fock 
parafermions in terms
of the multiplication of Fock states introduced above. So let
\begin{eqnarray}\label{cdagger}
C_i^\dagger|n_1,\dots,n_M\rangle
&\equiv&|n_i=1\rangle\times|n_1,\dots,n_M\rangle\\
&=&\bar{\omega}^{\sum_{j<i}n_j}\ |n_1,\dots,n_i+1,\dots, n_M\rangle\ .\nonumber
\end{eqnarray}
Because we know the inner product from Eq. \eqref{mb_inner}, we can immediately
compute the adjoint annihilation operators,
\begin{eqnarray}\label{c}
C_i|n_1,\dots,n_M\rangle
={\omega}^{\sum_{j<i} n_j}\ |n_1,\dots,n_i-1,\dots,n_M\rangle\ .
\end{eqnarray}
We can also define number operators as
\begin{equation}\label{basicn}
N_i|n_1,\dots,n_i,\dots,n_M\rangle=
n_i\, |n_1,\dots,n_i,\dots,n_M\rangle\ .
\end{equation}

The creation operators satisfy
\begin{eqnarray}\label{creation}
C_i^{\dagger p}=0\ , \qquad 
C_i^{\dagger }C_j^{\dagger }=\omega C_j^{\dagger }C_i^{\dagger }\quad (i<j)\ ,
\end{eqnarray}
and analogous relations follow for the annihilation operators, 
\begin{equation}\label{destruction}
C_i^{p}=0\ , \qquad C_iC_j=\omega C_jC_i \quad (i<j)\ . 
\end{equation}
Creation and annihilation operators for different orbitals commute up to 
a phase, 
\begin{eqnarray}\label{dc}
C_i^\dagger C_j=\bar{\omega}C_jC_i^\dagger\ , \qquad
C_iC_j^\dagger=\bar{\omega}C_j^\dagger C_i  \quad (i<j)\ .
\end{eqnarray}
But what are the relations for \(C\)s and \(C^\dagger\)s 
for a given orbital $i$? Canonical fermions 
require one relation per orbital, Eq. \eqref{cfrels}. 
In contrast, Fock parafermions require \(p-1\) relations,
\begin{equation}\label{wick}
C_i^{\dagger m}C_i^m+C_i^{p-m}C_i^{\dagger (p-m)}=\mathds{1}\qquad m=1,\cdots,p-1\ .
\end{equation}
This section's relations summarize the algebra of creation and annihilation 
operators of Fock parafermions. For \(p=2\), they reduce to the standard fermionic 
algebra. 

One of the landmark features of canonical fermions and bosons, and also some algebraic 
frameworks for Abelian anyons \cite{batista_ortiz}, is the simple relation
between the creation and annihilation operators and number operators. 
This relation is somewhat less simple for Fock parafermions since
\begin{equation}\label{compositen}
N_i=\sum_{m=1}^{p-1}C^{\dagger m}_iC^m_i.
\end{equation} 
However, it is still true that \(N_i\) is the generator of 
\(U(1)\) transformations. To see this, notice that
\begin{eqnarray}
[C^{\dagger m}_iC_i^m, C_i]= C^{\dagger m}_iC_i^{m+1}-C^{\dagger m-1}_iC_i^m
\end{eqnarray}
from Eq. \eqref{wick}. It follows that 
\begin{equation}\label{u1commutator}
[N_i,C_i^\dagger]=C_i^\dagger,\qquad  [N_i,C_i]=-C_i\ .
\end{equation}

Any operator in Fock space has a normal form \(:\mathcal{O}:\), defined implicitly as
\begin{equation}
\mathcal{O}=:\mathcal{O}:+\langle 0|\mathcal{O}|0\rangle,\qquad
:\mathcal{O}:|0\rangle=0=:\mathcal{O}:^\dagger|0\rangle\ . 
\end{equation}
Remarkably, the unconventional algebra of creation and annihilation operators for 
Fock parafermions provides a systematic way to compute \(:\mathcal{O}:\) by repeated 
application of Eqs. \eqref{creation}, \eqref{destruction}, \eqref{dc}, and \eqref{wick} 
to put all the creation operators to the left of every annihilation operator. But
for fermions (\(p=2\)) a more efficient procedure exists: 
the algorithm known as Wick's theorem
\cite{ballentine}. We do not know whether some generalization of Wick's theorem exists
that applies to Fock parafermions. 

Let us note in closing that if one takes the view that the orbitals
denote sites on a latttice (localized orbitals), 
Fock parafermions can be defined in any number of 
dimensions. This point is directly linked to the analogous discussion for 
parafermions around Eq. \eqref{uvs2pfsn}, and we will clarify it further  in \ref{whcb}.

\section{Fock representation of Weyl parafermions}
\label{cs_to_gs}

In this section we exploit the Fock parafermions just defined
to construct a Fock space representation of parafermions. This 
construction will generalize the representation of Majoranas in terms of 
ordinary fermions to all \(p\).  

Unfortunately, Eq. \eqref{m2f} as
it stands suggests the wrong starting point. Let
 \(\kappa\equiv e^{\im \pi/p}\), so that \(\kappa^2=\omega\).
Then the combination 
\begin{equation}\label{theta}
\chi_i\equiv\Gamma_{i}+\kappa\Delta_{i},
\end{equation}
satisfies \cite{kwasniewski,traubenberg}
\begin{eqnarray}
\chi_i^p=0,\qquad \qquad \chi_i\chi_j=\omega\chi_j\chi_i\quad (i<j).
\end{eqnarray}	
Moreover, for \(p=2\),
\(\chi_i=2C_i\) and \(\chi_i^\dagger=2C_i^\dagger\) are proportional
to fermionic annihilation and creation operators. This suggest that
we set \(C_i \propto \chi_i,\ C_i^\dagger \propto \chi_i^\dagger\), {\it
for all \(p\)}. But this ansatz fails to satisfy Eq. \eqref{wick}.
This, in our opinion, seems to be the reason why the algebra of creation 
and annihilation operators
described in this paper is new. The natural candidates for creation and 
annihilation operators associated to parafermions, 
the \(\chi_i\) of Eq. \eqref{theta} long known in the 
literature \cite{kwasniewski,traubenberg}, do not 
satisfy the correct algebra (in particular, they cannot be normal-ordered) 
to grant a particle interpretation.

The correct starting point can be obtained from rewriting 
Eq. \eqref{m2f} in the form  
\begin{equation}\label{m2fbetter}
\Gamma_{i}=a_i=C_i^{\;}+C_i^\dagger,\qquad 
\Delta_i=-\im b_i= (C_i^{\;}+C^\dagger_i)(-1)^{C_i^\dagger C_i^{\;}}\ ,
\end{equation}
which suggests a suitable generalization,
\begin{equation}\label{cs2gs}
\Gamma_{i}\equiv C_i^{\;}+C_i^{\dagger p-1}\ ,\qquad 
\Delta_i\equiv (C_i^{\;}+C_i^{\dagger p-1}) \, \omega^{N_i}\ ,
\end{equation}
valid for all \(p\). This is the Fock representation of parafermions
we have been looking for, and a main result of this paper.

The number operator \(N_i\) can be eliminated 
from Eq. \eqref{cs2gs} with the use of the identity
\begin{equation}\label{expand_phase}
\omega^{N_i}=
\mathds{1}+(\omega-1)\sum_{m=1}^{p-1}\omega^{m-1} C_i^{\dagger m}C_i^m,
\end{equation} 
that reduces to the well known \((-1)^{C_i^\dagger C_i^{\;}}=1-2C_i^\dagger C_i^{\;}\) for
ordinary fermions (\(p=2\)). 
Since, by Eq. \eqref{wick},
\begin{equation}
C_iC_i^{\dagger m}C_i^m=C_i^{\dagger m-1}C_i^m\ ,
\end{equation}
we have that
\begin{equation}
\Gamma_i=C_i+C_i^{\dagger p-1},\qquad 
\Delta_i=\omega C_i+ C_i^{\dagger p-1}
+(\omega-1)\sum_{m=2}^{p-1} C_i^{\dagger m-1}C_i^m\ .
\end{equation}
It is possible to invert Eq. \eqref{cs2gs} to obtain an expansion 
of creation and annihilation operators for Fock parafermions 
in terms of parafermions. This expansion follows most easily from
results obtained in the \ref{whcb}, so here we just quote the expressions,
\begin{eqnarray}
C_i&=&\frac{p-1}{p} \, \Gamma_i-
\frac{1}{p}\sum_{m=1}^{p-1}\omega^{m(m+1)/2} \, \Gamma_i^{m+1}\Delta_i^{\dagger m},\nonumber \\
C_i^\dagger&=&\frac{p-1}{p} \, \Gamma_i^\dagger-
\frac{1}{p}\sum_{m=1}^{p-1}\bar{\omega}^{m(m+1)/2} \, \Delta_i^{m}(\Gamma_i^\dagger)^{m+1}
\ .
\label{gs2cs}
\end{eqnarray}

Finally, let us exploit Eqs. \eqref{parafermionic_vp} and \eqref{cs2gs} to reinterpret 
the vector Potts model as the Hamiltonian for a system of Fock parafermions ($J_M=0$),
\begin{eqnarray}\label{vpweyl}
H_{\sf VP}=-\frac{1}{2}
\Big[\sum_{i=1}^M h_i \, \bar{\omega}^{N_i}+\sum_{i=1}^{M-1} J_i 
(C_i^{\;}C^\dagger_{i+1}+C_i^{\dagger p-1}C_{i+1}^{p-1})\, \omega^{N_{i}}
\label{u1symclock}\\
+\sum_{i=1}^{M-1} J_i (C_i^{\;}C_{i+1}^{p-1}+C_i^{\dagger p-1}C_{i+1}^\dagger) \,
\omega^{N_{i}}  \Big]+{\sf h.c.}\ .\label{anomalous}
\end{eqnarray}
From this point of view, the vector Potts Hamiltonian splits into two 
terms, Eqs. \eqref{u1symclock} and \eqref{anomalous}, 
with different symmetry properties. The first term, Eq. \eqref{u1symclock}, is 
\(U(1)\) symmetric because it commutes with the operator
\(\hat{N}=\sum_{i=1}^M N_i\) counting the total number of Fock parafermions. 
The second type of term, Eq. \eqref{anomalous}, breaks this continous symmetry down to 
a discrete \(\mathds{Z}_p\) symmetry \(\omega^{\hat{N}}\) via anomalous pairing. 
The existence  of an emergent \(U(1)\) symmetry in the vector Potts model and its
exact nature was a subject of discussion for many years, see Ref. \cite{pclock}
and references therein. From the present analysis  we can understand these features
in terms of particle conservation broken down, due to anomalous pairing, to a 
discrete symmetry.   


\vspace*{0.7cm}
\noindent
{\bf Acknowledgements}

EC dedicates this paper to the memory of China Nieves Corriente de and 
Alberto Glavocich, and gratefully acknowledges support from 
the Dutch Science Foundation NWO/FOM and an ERC Advanced Investigator Grant.

\appendix

\section{Weyl hard-core bosons}
\label{whcb}

In the main body of the paper we established ``dictionaries" \cite{batista_ortiz}
connecting Weyl generators to parafermions, Eqs.  \eqref{uvs2pfs} and \eqref{gd2uvs},
and Fock parafer\-mions to parafermions, Eqs. \eqref{cs2gs} and \eqref{gs2cs}. 
In this appendix we would like to combine some of these results to obtain a Fock
representation of  Weyl generators that helps in the derivation of  Eq. \eqref{gs2cs}. 
A simple calculation reveals a new Fock space of particles
satisfying bosonic exchange statistics and \(p\)-exclusion.
We call these particles {\it Weyl hard-core bosons}. The latter
provide a local Fock space description
of Weyl generators, just like Fock parafermions are best suited for the local 
Fock space description of parafermions. The creation and 
annihilation operators for Weyl hard-core bosons
are connected to those of Fock parafermions by a Jordan-Wigner-like
transformation. 

Consider expressing the Weyl generators in terms of parafermions. From 
Eqs. \eqref{gd2uvs},  and \eqref{cs2gs} 
\begin{eqnarray}
U_i=\Gamma_i^\dagger\Delta_i^{\;}=
(C_i^\dagger+C_i^{p-1})(C_i^{\;}+C_i^{\dagger p-1})\, \omega^{N_i}=\omega^{N_i}\ ,
\end{eqnarray}
which follows from \(p\)-exclusion
and Eq. \eqref{wick}, and  ($(\omega^{N_i})^\dagger=\bar{\omega}^{N_i}$)
\begin{eqnarray}
V_i \!\!\! &=& \!\!\! \Gamma_i^{\;}\Big( \prod_{j=1}^{i-1} 
\Delta_j^\dagger \Gamma_j^{\;}\Big)=
(C_i^{\;}+C_i^{\dagger p-1})\Big(\prod_{j=1}^{i-1} \bar{\omega}^{N_j} \! \Big)
=C_i\prod_{j=1}^{i-1} \bar{\omega}^{N_j} \! + \!
\Big( \! C_i^\dagger\prod_{j=1}^{i-1} \omega^{N_j} \! \Big)^{p-1}
\nonumber\\ 
&=&B_i+B_i^{\dagger p-1} \ , 
\end{eqnarray}
where we have defined creation and annihilation operators
\begin{equation}\label{ujw}
B_i^\dagger=C_i^\dagger\, \prod_{j=1}^{i-1}\omega^{N_j},\qquad 
B_i=C_i\, \prod_{j=1}^{i-1}{\bar \omega}^{N_j}\ .
\end{equation}
Noticing that 
\begin{eqnarray}\label{nwhcb}
N_i=\sum_{m=1}^{p-1} C_i^{\dagger m}C_i^m=\sum_{m=1}^{p-1} B_i^{\dagger m}B_i^m\ ,
\end{eqnarray}
we thus managed to express the Weyl generators in terms of  \(B_i^{\;}$, and $B_i^\dagger\).

The key point is that these operators satisfy 
bosonic-type commutation relations
\begin{equation}\label{they_commute}
[B_i,B_j]=0,\qquad [B_i^{\;},B_j^\dagger]=0,\qquad [B_i^\dagger,B_j^\dagger]=0.
\end{equation}
Furthermore,
\begin{equation}\label{lower_rel}
B_i^p=0=B_i^{\dagger p},\qquad B_i^{\dagger m}B_i^m+B_i^{p-m}B_i^{\dagger (p-m)}=
\mathds{1}\quad (m=1,\cdots,p-1)\ ,
\end{equation}
with number operators, \(N_i=\sum_{m=1}^{p-1} B_i^{\dagger m}B_i^m\), 
which are generators of \(U(1)\) transformations,
\begin{equation}
[N_i,B_i]=-B_i,\qquad [N_i^{\;}, B_i^\dagger]= B_i^\dagger.
\end{equation}
These creation and annihilation operators are the 
Weyl hard-core bosons alluded to at the beginning of this appendix. 
Just as for Fock parafermions, any operator polynomial
in the \(B_i^\dagger, B_i^{\;}\) can be put in normal form just by repeated application
of the defining relations \eqref{they_commute} and \eqref{lower_rel}. 

Equations \eqref{they_commute} and \eqref{lower_rel} imply that 
the Fock space of Weyl hard-core bosons 
describes bosonic \(\theta=0\) exchange and \(p\)-exclusion statistics,
\begin{eqnarray}
|n_i\rangle\times|n_j\rangle= |n_j\rangle \times |n_i\rangle=|n_i,n_j\rangle, 
\quad (|n_i=1\rangle)^p=0\ .
\end{eqnarray}
Notice that in this case the exclusion and exchange statistics are independent.
This is clear since the latter depends on \(p\) while the former is fixed.

Weyl hard-core bosons can be expanded directly in terms of Weyl
generators. Since
\begin{equation}
\langle n_i=m|V_i^\dagger B_i^{\;} |n_i=n\rangle=\left\{
\begin{array}{lcl}
0 & \mbox{if} & n\neq m\\
0 & \mbox{if} & n=m=0 \\
1 & \mbox{if} & n=m=1,\dots, p-1
\end{array} \right .\ ,
\end{equation}
the diagonal operator \( V_i^\dagger B_i^{\;}\) can be expanded
as \( V_i^\dagger B_i^{\;}=\sum_{m=0}^{p-1} c_m U_i^{\dagger m}\),
with expansion coefficients  \(c_m={\sf tr}(U_i^mV_i^\dagger B_i^{\;})/p\). 
It follows that 
\begin{equation}\label{uvs2as}
B_i=\frac{p-1}{p}V_i-\frac{1}{p}\sum_{m=1}^{p-1}V_iU_i^{\dagger m},\qquad
B_i^\dagger=\frac{p-1}{p}V_i^\dagger-\frac{1}{p}\sum_{m=1}^{p-1}U_i^{ m}V_i^\dagger,
\end{equation}
or in matrix representation, as follows from Eq. \eqref{VsandUs}, 
\begin{eqnarray} \hspace*{-0.5cm}
{B}=
\begin{pmatrix}
0& 1& 0& \cdots& 0\\
0& 0& 1& \cdots& 0\\
0& 0& 0& \cdots& 0\\
\vdots& \vdots& \vdots&      & \vdots \\
0& 0& 0& \cdots& 1\\
0& 0& 0& \cdots& 0\\
\end{pmatrix}, \qquad
{B^\dagger}=
\begin{pmatrix}
0& 0& 0& \cdots& 0& 0\\
1& 0& 0& \cdots& 0& 0\\
0& 1& 0& \cdots& 0& 0\\
\vdots & \vdots & \vdots & &\vdots & \vdots\\
0& 0& 0& \cdots& 1& 0
\end{pmatrix}
\ .  \label{matrixbs}
\end{eqnarray}
Notice that 
\begin{equation}
B_iU_i=\omega U_iB_i,\qquad U_iB_i^\dagger=\omega B_i^\dagger U_i\ .
\end{equation}

We have now the elements to derive Eq. \eqref{gs2cs} of Section \ref{cs_to_gs}. 
By combining Eqs. \eqref{uvs2as} and \eqref{ujw},  and using the definition of parafermions 
Eq. \eqref{uvs2pfsn}, we obtain Eq. \eqref{gs2cs}.
Fock parafermions can be defined {\it in any number of space dimensions}
by  exploiting the 
generalized Jordan-Wigner transformation
\begin{equation}\label{hdimujw}
C_\r^\dagger=B_\r^\dagger \prod_{\x<\r} U_\x^\dagger,\qquad 
C_\r=B_\r \prod_{\x<\r} U_\x ,
\end{equation}
with 
\begin{equation}
U_\x=\omega^{N_\x}=\mathds{1}+(\omega-1)\sum_{m=1}^{p-1}\omega^{m-1}B_\x^{\dagger m}B_\x^m.
\end{equation}
Weyl hard-core bosons become the standard hard-core bosons \cite{batista_ortiz}
($U_\x=\sigma^z_\x=1-2N_\x$) for $p=2$, and 
Eq. \eqref{hdimujw} reduces to the Matsubara-Matsuda transformation in any 
number of dimensions. 

We finally 
describe the mathematical connection between Weyl hard-core bosons and an alternative
description of hard-core bosons, the so-called {\it g-ons} introduced in Ref. \cite{batista_ortiz},  
and later on used 
in connection to the simulation of competition and coexistence of magnetism and 
superfluid behaviors in cold atoms and optical lattice systems \cite{hmft}. Define
\begin{equation}
g_i\equiv B_i^{\;} \, N_i^{1/2},\qquad g_i^\dagger\equiv  N_i^{1/2} \, B_i^\dagger\ ,
\end{equation}
and let \(W_i=W_i^2=W_i^\dagger\) denote the orthogonal projector onto
the subspace of states with maximal occupation \(p-1\) of orbital \(i\).
The operators \(g_i^{\;},g_i^\dagger, W_i^{\;}\ (i=1,\dots,M\)) offer an alternative,
{\it yet not a Lie algebra}, description of  hard-core
bosons with $p$ exclusion that converges naturally to canonical bosons in the limit 
\(p\rightarrow\infty\), and to standard hard-core bosons for $p=2$. 
Their algebra is completely specified by the relations
\begin{equation}
[g_i^{\;},g_j^{\;}]=0=[g_i^{\dagger},g_j^{\dagger}] , \qquad
[g_i^{\;},g_j^\dagger]=\delta_{i,j}(\mathds{1}-p \, W_i),\qquad 
g_i^\dagger \, W_i^{\;}=0=W_i \, g_i\ ,
\end{equation}
and \(p\)-exclusion \cite{batista_ortiz}. For any $p$, $N_i=g_i^\dagger g_i^{\;}$. One may 
argue that if the weak limit 
\begin{equation}
w-\lim_{p\rightarrow\infty} W_i = 0
\end{equation}
holds, then $g_i^{\;},g_i^\dagger$ converge to canonical bosons in 
the limit \(p\rightarrow\infty\). 
A weak limit is established through the convergence of matrix elements.
The matrix elements of \(W_i\) converge to zero  because, by construction,
there are no physical states with infinite occupation number in Fock space.  

Let us emphasize that {\it g-ons} with \(p\)-exclusion 
are best suited for the local Fock space description of spins of dimension
\(2S+1=p\). In contrast, an attempt to describe Weyl generators in terms of 
{\it g-ons} would 
return hopelesly complicated expressions (especially for  relatively large 
\(p\)). Conversely,  an attempt to describe spins in terms of Weyl hard-core bosons 
would again return complicated expressions (especially for  relatively large 
\(S\)).



\begin{thebibliography}{}
\bibitem{norm} 
The reason is remarkably simple. It follows from Eq. \eqref{heisenberg} 
that \([X_i^n,P^{\;}_i]=\im \hbar nX_i^{n-1}\ (n=2,3,\dots)\). But then a representation
of Heisenberg's commutation relations in terms of bounded operators 
would have to satisfy the inequalities  
\[ n\hbar\leq 2 \|X_i\| \cdot \|P_i\|,\qquad n=2,3,\dots\ ,\]
which clearly contradicts the assumption that both \(X_i\) and \(P_i\) are bounded.
In particular, there is no finite-dimensional representation of Heisenberg's relations.
\bibitem{q_compu}
The quantum control model of quantum computation assumes
the existence of physical systems that can be controlled
by modulating the parameters of the system's Hamiltonian. 
The control possibilities are used to implement specific 
quantum gates that represent the unitary
evolution of the physical system over a time step obtained
by specific modulations of the Hamiltonian. 
In this regard, we can think of the clock Hamiltonian $H_{\sf VP}$
of Eq. \eqref{pclockH} as our system Hamiltonian whose parameters 
$\{h_i,J_i\}$ are controlled. 
The quantum control and quantum gate viewpoints are effectively
equivalent \cite{fermionic_comp}.
\end{thebibliography}
\end{document}